\def\apj{{ApJ}}
\def\apjs{{ApJS}}

\catcode`\@=11
\def\gsim{\ifmmode{\mathrel{\mathpalette\@versim>}}
    \else{$\mathrel{\mathpalette\@versim>}$}\fi}
\def\lsim{\ifmmode{\mathrel{\mathpalette\@versim<}}
    \else{$\mathrel{\mathpalette\@versim<}$}\fi}
\def\@versim#1#2{\lower 2.9truept \vbox{\baselineskip 0pt \lineskip
    0.5truept \ialign{$\m@th#1\hfil##\hfil$\crcr#2\crcr\sim\crcr}}}
\catcode`\@=12
\def\msun{\hbox{$M_\odot$}}
\def\lsun{\hbox{$L_\odot$}}

\def\zfe{\hbox{$Z^{\rm Fe}$}}
\def\zfesun{\hbox{$Z_\odot^{\rm Fe}$}}

\def\lb{\hbox{$L_{\rm B}$}}

\def\msun{\hbox{$M_\odot$}}
\def\m*{\hbox{$M_*$}}
\def\IMLR{$M^{\rm Fe}/L$}
\def\micm{\hbox{$M_{\rm ICM}$}}
\def\mfecm{\hbox{$M^{\rm Fe}_{\rm ICM}$}}

\def\zfe{\hbox{$Z^{\rm Fe}$}}
\def\zfes{\hbox{$Z^{\rm Fe}_{*}$}}
\def\zfecm{\hbox{$Z^{\rm Fe}_{\rm ICM}$}}
\def\ho{\hbox{$H_\circ$}}
\def\h50{\hbox{$\ho /50$}}
\def\h70{\hbox{$h_{70}$}}
\def\yr-1{\hbox{${\rm yr}^{-1}$}}

\documentclass[cup5b]{caps}
\usepackage{graphicx}
\usepackage{amssymb}
\usepackage{ociwsymp3}   

\HeadText{A. Renzini}  

\begin{document}

\pagenumbering{arabic}

\author[]{ALVIO RENZINI \\ European Southern Observatory}

\chapter{The Chemistry of Galaxy Clusters}

\begin{abstract}

From X-ray observations of galaxy clusters one derives the mass of the
intracluster medium along with its chemical
composition. Optical/infrared observations are used to estimate the mass of
the stellar components of galaxies, along with their chemical
composition and age. This review shows that when combining all this
information, several interesting inferences can be drawn, including:
(1) galaxies lose more metals than they retain; (2) clusters and the
general field have converted the same fraction of baryons into stars,
hence the metallicity of the $z = 0$ Universe as a whole has to be
nearly the same we see in clusters, $\sim$1/3 solar; (3) for the same
reason, the thermal content of the intergalactic medium is expected to
be nearly the same as the preheating energy of clusters; (4) a strong
increase of the Type Ia supernova (SN) rate with lookback time is predicted 
if SNe~Ia produce a major fraction of cosmic iron; (5) the global metallicity of
the $z\approx 3$ Universe was already $\sim$1/10 solar; and (6) the Milky Way
disk formed out of material that was pre-enriched to $\sim$1/10 solar
by the bulge stellar population.

\end{abstract}

\section{Introduction}

In one of the rare cases in which theory anticipates observations, the
existence of large amounts of heavy elements in the intracluster
medium (ICM) was predicted shortly before it was actually observed (Larson \& 
Dinerstein 1975). This
came from (now old-fashioned) so-called {\it monolithic} models of
elliptical galaxy formation, in which the observed color-magnitude
relation is reproduced in terms of a metallicity trend. In turn, this
trend is established by SN-driven galactic winds being more
effective in less massive, fainter galaxies with shallow potential wells,
compared to more massive galaxies harbored in deep potential wells.
While these models may now be inadequate in quite many respects, 
their prediction was confirmed the following year by the discovery of the 
strong iron-K line in the X-ray spectrum of galaxy clusters (Mitchell et al. 
1976).

This Carnegie Symposium on clusters of galaxies covers all the
manifold aspects of these largest bound systems in the Universe. This
review is meant to focus on one specific topic, the metal content of
the clusters. I show that from this one can infer quite a number
of intriguing consequences on galaxy formation and evolution on a wide
scale, as well as on the evolution of some global properties of the
baryonic component of the Universe. The following will be a {\it
broad-brush} picture about facts and inferences, and is meant to
stimulate a deeper look at each of the issues that will be cursorily
touched upon here, and which  include the following main topics:

\begin{itemize}

\item
The metal content of clusters: ICM and galaxies

\item
The composition of the ICM  ``metallicity'' (elemental ratios)

\item
The ICM/galaxies iron share in clusters

\item
Metal production: Type Ia and Type II SNe, and the cosmic evolution of their 
rate

\item
Metal transfer from galaxies to ICM: ejection vs. extraction

\item
Metals as tracers of the ``ICM preheating''

\item
Clusters vs. field at $z=0$

\item
The major epoch of metal production in clusters

\item
The metallicity of the Universe at  $z=3$

\item
The early chemical evolution of the Milky Way

\end{itemize}

The production and circulation of iron and other heavy elements on galaxy 
cluster scale has been widely discussed since their early discovery
(e.g.,  Vigroux 1977; Matteucci \& Vettolani 1988; Ciotti et al. 1991; Arnaud 
et al. 1992; Renzini et al. 1993; Loewenstein \& Mushotzky 1996; Ishimaru \& 
Arimoto 1997; Renzini 1997, 2000; Chiosi 2000; Aguirre et al.  2001; Pipino et 
al. 2002). 

\section{The Heavy Elements in Clusters: ICM and Galaxies}

\subsection{Iron}
Iron is the best studied element in clusters of galaxies, as ICM iron emission 
lines are present in all clusters and groups, either warm or hot.
Figure 1.1 shows the iron abundance in the ICM of clusters and groups as a
function of ICM temperature from an earlier compilation (Renzini 
2000).  For $kT\gsim 3$ keV the ICM iron abundance is constant at
$\zfe\simeq 0.3 \zfesun$, independent of cluster
temperature. Abundances for clusters in this {\it horizontal} sequence
come from the iron-K complex at $\sim 7$ keV, whose emission
is due to transitions to the K level of H-like and He-like iron ions.
At lower temperatures the situation is much less simple. Figure 1.1
shows data from Buote (2000), with the iron abundance having been
derived with both one-temperature and two-temperature fits. The
one-temperature fits give iron
abundances for those cool groups that are more or less in line with those of 
the hotter clusters. The abundances of the two-temperature fits, instead, 
form an almost vertical sequence, with
a great deal of dispersion around a mean value of $\sim 0.75$ solar.
Earlier estimates gave extremely low values for cooler groups, $kT\lsim 1$ keV
(Mulchaey et al. 1996). Compiling values from the literature, a strong 
dependence of the abundance on ICM temperature is apparent, being very low at
low temperatures, steeply increasing to a maximum around $kT \approx 2$ keV,
then decreasing to reach $\sim 0.3$ solar by $kT \approx 3$ keV (Renzini 1997;
see also Mushotzky 2002).

\begin{figure*}[t]
\centering
\includegraphics[width=1.00\columnwidth,angle=0,clip]{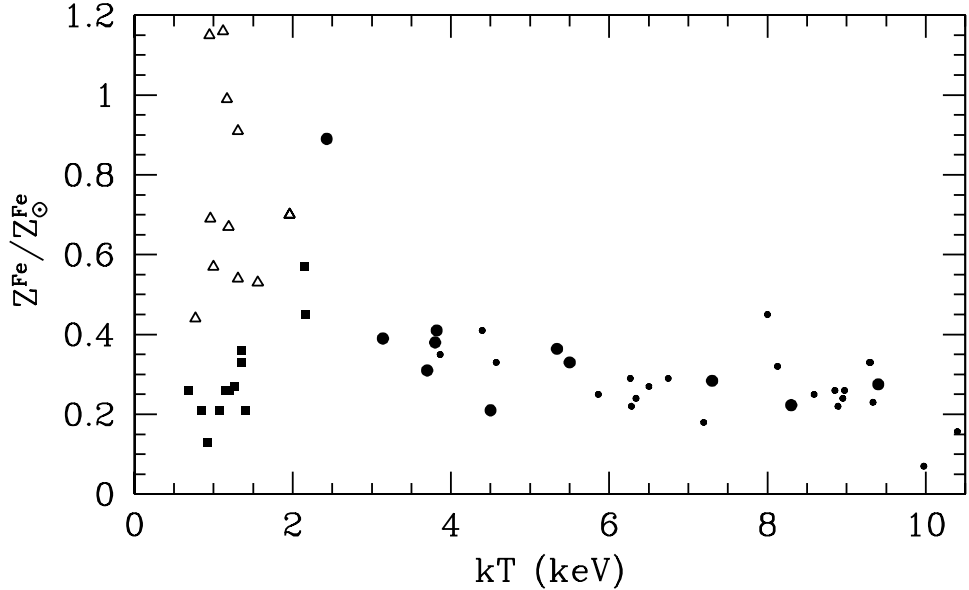}
\vskip 0pt \caption{
A compilation of the iron abundance in the ICM as a function of the ICM 
temperature for a sample of clusters and groups (Renzini 2000), including 
several clusters at moderately high redshift with 
$\langle z\rangle \simeq 0.35$, represented by small filled circles. For 
temperatures less than about 2 keV, 11 groups are shown from Buote (2000), 
with temperatures and abundances determined from one- and two-temperature fits 
(filled squares and open triangles, respectively).}
\end{figure*}

\begin{figure*}[t]
\centering
\includegraphics[width=1.00\columnwidth,angle=0,clip]{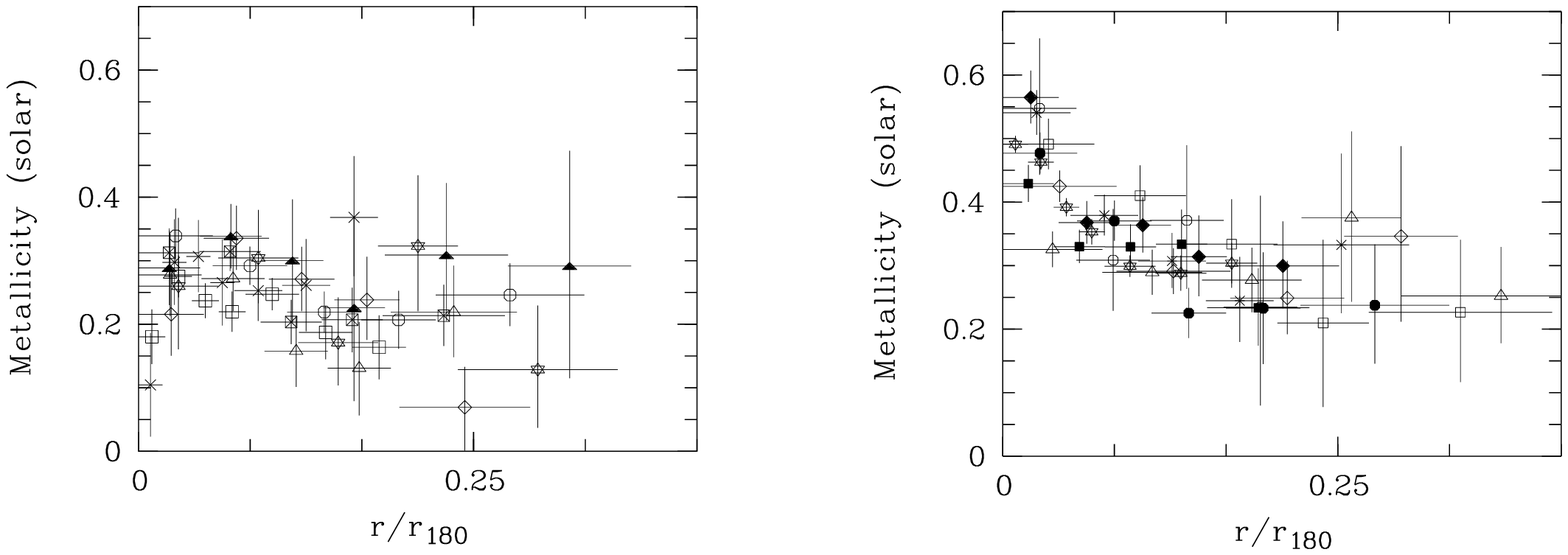}
\vskip 0pt \caption{
Projected metallicity distributions for non-cold core clusters (left panel)
and cold-core clusters (right panel), from {\it Beppo-SAX} data (De~Grandi
\& Molendi 2001). The radial coordinate is normalized to the radius
with  an overdensity factor of 180.}
\end{figure*}

Is this strong temperature dependence real? Perhaps some caution
is in order. Besides the ambiguity as to whether one- or 
two-temperature fits are preferable, 
additional uncertainties for the iron abundances at
$kT\lsim 2$ keV come from their being derived from the iron-L
complex at $\sim 1$ keV, whose emission lines are due to transitions
to the L level of iron ions with three or more electrons. In these
cooler groups/clusters iron is indeed in such lower ionization stages, and
the iron-K emission disappears. The atomic configurations of these more complex
ions are not as simple as those giving rise to the iron-K emission, 
and their (calculated) collisional excitation probabilities may be more 
uncertain. In summary, iron abundances derived from the iron-L emission
should be regarded with a little more caution compared to those from the
iron-K emission.

 Abundances shown in Figure 1.1 refer to the cluster central regions.
 However, radial gradients in the iron abundance have been reported
 for several clusters, starting with {\it ASCA} and then {\it ROSAT} data
 (e.g., Fukazawa et al. 1994; Dupke \& White 2000; Finoguenov, David, 
 \& Ponman 2000; White 2000; Finoguenov, Arnaud, \& David 2001).
 From {\it Beppo-SAX} data, De~Grandi \&
 Molendi (2001) have conducted a systematic study of the radial
 distribution of iron (metals) in many clusters. Figure 1.2 shows that
 clusters break up into two distinct groups: so-called cool core 
 clusters (formerly known as ``cooling flow'' clusters before
 the failure of the cooling flow model was generally acknowledged) are
 characterized by a steep metallicity (mostly iron) gradient in the
 core, reaching $\sim 0.6$ solar near the center, and non-cold core
 clusters (where no temperature gradient is found), which show no metallicity
 gradient.  The origin of the dichotomy remains to be understood.  The fact 
 that metallicity gradients are found to be associated with
 large temperature gradients in the central regions may
 look suspicious, as noted for the strong dependence of $\zfe$ on ICM
 temperature, but it appears to be well established.

\subsection{Elemental Ratios}

X-ray observatories (especially {\it ASCA}, {\it Beppo-SAX}, and {\it XMM-Newton}) have
such high spectral resolution that besides those of iron the emission
lines of many other elements can be detected and measured. These
include oxygen, neon, magnesium, calcium, silicon, sulfur, argon, and
nickel. Most of these are $\alpha$ elements, predominantly synthesized
in massive stars exploding as Type II SNe. As is well known,
iron-peak elements are mainly produced by Type Ia SNe, and
$50\%-75\%$ of the iron in the Sun may come from them.

Early estimates from {\it ASCA} suggested a sizable $\alpha$-element
enhancement, $\langle$[$\alpha$/Fe]$\rangle$ $\simeq$ +0.4 (Mushotzky (1994), 
later reduced to +0.2 (Mushotzky et al. 1996), and eventually found
consistent with solar proportions $\langle$[$\alpha$/Fe]$\rangle$ $\simeq$ 0.0
(Ishimaru \& Arimoto 1997). More recently, Finoguenov et al. (2000)
report near-solar Ne/Fe, slightly enhanced Si/Fe, and slightly
depleted S/Fe, but with rather large error bars.  From a systematic
reanalysis of the {\it ASCA} archival data, Mushotzky (2002) reports a
systematic increase of Si and Ni and a decrease of Ca with ICM
temperature. Note that both silicon and calcium are $\alpha$ elements,
and apparently they do not follow the same trend! 
No simple interpretation has so far emerged of these  trends in terms
of the relative role of the two SN types (Gibson \& Matteucci
1997; Loewenstein 2001; Finoguenov et al. 2002). 
 
I would conclude that no compelling evidence exist for other than near-solar 
[$\alpha$/Fe] ratios in the ICM, when all $\alpha$ elements are
lumped together. This argues for stellar nucleosynthesis having
proceeded in much the same way in the solar neighborhood as well as at
the galaxy cluster scale.  In turn, this demands a similar ratio of
the number of Type Ia to Type II SNe, as well as a similar stellar initial 
mass function (IMF), suggesting that the star formation process (IMF, binary 
fraction, etc.) is universal, with little or no dependence on the global
characteristics of the parent galaxies (and their large-scale
structure environment) in which molecular clouds are turned into
stars.  Alternatively, one can take at face value the variations of
the abundance ratios with cluster temperature, as well as the
overabundance of some $\alpha$ elements and the underabundance of
others.  One can then be forced to rather contrived conclusions, such
as the mix of the two SN types, and perhaps even the nucleosynthesis
of massive stars, depends on what the temperature of the ICM will be
billions of years {\it after} star formation has ceased. On the other
hand, one may argue that rich galaxy clusters are ``special'' places
in many senses, and that ICM abundances reflect not only SN
nucleosynthesis yields, but also how efficiently these are ejected,
mixed into, and retained in the ICM. However, no simple understanding
of the apparent empirical trends has yet emerged.

In summary, in the following I will assume that clusters, on a global scale,
have solar elemental ratios and the total heavy element abundance is 0.3 solar,
or 0.006 by mass.

\subsection{The Iron Mass-to-Light Ratio}

One useful quantity is the iron-mass-to-light-ratio (\IMLR) of the
ICM, the ratio $\mfecm/\lb$ of the total iron mass in the ICM
over the total $B$-band luminosity of the galaxies in the cluster. In
turn, the total iron mass in the ICM is given by the product of the
iron abundance times the mass of the ICM, 
$\mfecm=\micm\zfecm$. Figure 1.3 shows the resulting \IMLR \ from an
earlier compilation (Renzini 1997).  The drop of the \IMLR \ in poor
clusters and groups (i.e., for $kT\lsim 2$ keV) can be traced back to a
drop in both the iron abundance (which, however, may not be real; see
above) {\it and} in the ICM mass. Such groups appear to be gas poor
compared to clusters, which suggests (1) that they may have been subject
to baryon and metal losses due to strong galactic winds driving much
of the ICM out of them (Renzini et al. 1993; Renzini 1997; Davis, Mulchaey, 
\& Mushotzky 1999), (2) that such winds have {\it preheated} the gas around
galaxies, thus preventing it to fall inside groups, or (3) that they have
{\it inflated} the gas distribution. In one way or another, the {\it
break} seen in Figure 1.3 is likely to be related to the break of
self-similarity in the X-ray-temperature relation (see later).

For the rest of this paper I will mainly deal with clusters with
$kT\gsim 2-3$ keV, for which the interpretation of the data appears
more secure. Yet, several cautionary remarks are in order.  The first
is that the iron abundances used to construct Figure 1.3 did not take into
account that some clusters  have sizable iron gradients.  In
principle, X-ray observations can give both the run of gas density and
abundance with radius, so to make possible to integrate their product
over the cluster volume to get $\mfecm$. To my knowledge, so far this
has been attempted only for one cluster (Pratt \& Arnaud
2003). However, the extra iron contained within the core's iron gradient 
seems to be the product of the central cD galaxy, and may represent
only a small fraction of the whole $\mfecm$ (De~Grandi \& Molendi 2002).

Another concern is that two of the three ingredients entering into the
calculation of the \IMLR \ values shown in Figure 1.3 (namely $\micm$ and
$\lb$) may not be measured in precisely the same way in the various
sources used in the compilation. Both quantities come from a radial
integration up to an ill-defined cluster boundary, such as the Abell
radius, the virial radius, or to a radius of some fixed
overdensity. Sometimes it is quite difficult to ascertain what
definition has been used by one author or another, with the
complication that in general X-ray and optical data have been
collected by different groups using different assumptions. There is
certainly room for improvement here. Finally, estimated total
luminosities ($\lb$) refer to the sum over all cluster galaxies and do
not include the population of stars that is diffusely distributed throughout 
the cluster, which may account for at least $\sim 10\%$ of the total
cluster light (Ferguson, Tanvir, \& von Hippel 1998; Arnaboldi et al.  2003).

\begin{figure*}[t]
\centering
\includegraphics[width=1.00\columnwidth,angle=0,clip]{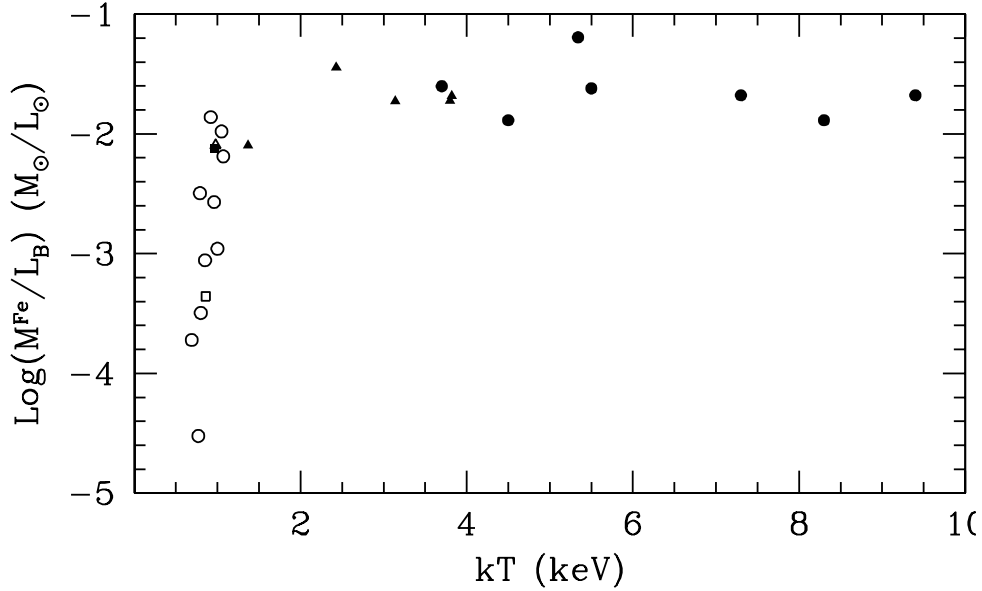}
\vskip 0pt \caption{
The iron mass-to-light ratio  of  the  ICM of clusters and groups as a 
function of the ICM temperature from an earlier compilation (Renzini 1997).
Data are taken from the following sources: filled circles, Arnaud
et al. (1992); filled triangles, Tsuru (1992); open triangle, David et al.
(1994); open square, Mulchaey et al. (1993); filled square,
Ponman et al. (1994); and open circles, Mulchaey et al. (1996).}
\end{figure*}

While keeping these cautions in mind, we see from Figure 1.3 that 
\IMLR \ runs remarkably flat with increasing cluster temperature, for
$kT\gsim 2-3$ keV. This constancy of the \IMLR \ comes from both
$\zfecm$ and $\micm/\lb$, showing very little trend with cluster
temperature [see Fig. 1.1 and Fig. 4 in Renzini (1997), where
$\micm/\lb\simeq 25h_{70}^{-1/2}\; (\msun/\lsun)$.] The resulting \IMLR \
is therefore 

\begin{equation}
{{M^{\rm Fe}_{\rm ICM}}\over{L}} =\zfecm{\micm\over\lb}\simeq 0.3\times\zfesun\times 25
     h_{70}^{-1/2} \simeq 0.01\,h_{70}^{-1/2} (\msun/\lsun).
\end{equation}
That is, in the ICM there is about 0.01 solar
masses of iron for each solar luminosity of the cluster galaxies. This
value is $\sim 30\%$ lower than adopted in Renzini (1997) and shown in
Figure 1.3, having consistently adopted here for $\zfesun$ the
recommended {\it meteoritic} iron abundance (Anders \& Grevesse 1989),
$\zfesun=0.0013$. Assuming solar elemental proportions for the ICM,
the ICM {\it metal mass-to-light ratio} is therefore $\sim 0.3\times
0.02\times 25 h_{70}^{-1/2} = 0.15\; (\msun/\lsun)$, having adopted $Z_\odot =
0.02$ and for $h_{70}=1$.

A very accurate analysis was performed recently for the A1983 cluster 
(Pratt \& Arnaud 2003), paying attention to measure $\micm$ and $\lb$
within the same radius. The result is  \IMLR$=(7.5\pm 1.5)\times 10^{-3}
h_{70}^{-1/2} (\msun/\lsun)$, in fair agreement with the estimate above.

The most straightforward interpretation of the constant \IMLR\ is that
clusters did not lose iron (hence baryons), nor differentially
acquired pristine baryonic material, and that the conversion of
baryonic gas into stars and galaxies has proceeded with the same
efficiency and the same stellar IMF in all clusters (Renzini 1997).
Otherwise, there should be  cluster-to-cluster variations of $\zfecm$
and \IMLR. All this is true insofar as the baryon-to-dark matter ratio is 
the same in all $kT\gsim 2$ keV clusters (White
at al. 1993), and the ICM mass-to-light ratio and the gas fraction are
constant. Nevertheless, there may be hints for some of these
quantities showing (small) cluster-to-cluster variations (Arnaud \& Evrard 1999;
Mohr, Mathiesen, \& Evrard 1999; Pratt \& Arnaud 2003), but no firm
conclusion has been reached yet.

\subsection{The Iron Share Between ICM and Cluster Galaxies}

The metal abundance of the stellar component of cluster galaxies is
derived from integrated spectra coupled to synthetic stellar
populations. Much of the stellar mass in clusters is confined to
passively evolving spheroids (ellipticals and bulges), for which the
iron abundance $\zfes$ may range from $\sim$1/3 solar to a few times solar.
For example, among ellipticals metal-sensitive spectral features such
as the magnesium index Mg$_2$ range from values slightly lower
than in the most metal-rich globular clusters of the Milky Way bulge
(which are nearly solar), to values for which models indicate a metallicity
a few times solar (e.g., Maraston et al. 2003). 
The \IMLR \ of cluster galaxies is then given by:

\begin{equation}
{{M^{\rm Fe}_{\rm gal}}\over{L}}=\zfes{\m*\over\lb}\simeq 0.0046\,h_{70}\quad 
(\msun/\lsun),
\end {equation}
where we have adopted $M_*/\lb=3.5\,h_{70}$ (White et al. 1993) and $\zfes=
\zfesun$.
The {\it total} cluster \IMLR \ (ICM+galaxies) is
therefore $\sim 0.015\quad (\msun/\lsun)$, for $h_{70}=1$, and
the ratio of the iron mass
in the ICM to the iron mass locked into stars and galaxies is

\begin{equation}
{\zfecm\micm\over\zfes M_*}\simeq 2.2 h_{70}^{-3/2},
\end{equation}
having adopted $\zfecm = 0.3$ $\zfesun$, $\zfes =1$ $\zfesun$, and $\micm
/M_*= 9.3h_{70}^{-3/2}$ as for the Coma cluster (White et al. 1993).
So, it appears that there is $\sim 2$ times more iron mass in the ICM than
locked into cluster stars (galaxies), perhaps even more if  $\zfes$ is subsolar
due to an abundance gradient within clusters (Arimoto et al. 1997). 
In turn, this empirical iron share (ICM vs.
galaxies) sets a strong constraint on models of the chemical evolution of
galaxies. Under the same assumptions as above, the
total metal mass-to-light ratio (ICM + galaxies) is therefore $\sim  
0.15h_{70}^{-1/2}+0.07h_{70}\simeq 0.2\;  (\msun/\lsun)$. This can be regarded
as a fully empirical determination of the metal yield of (now) old stellar
populations.

\section{Metal Production: The Parent Stellar Population}

The constant \IMLR \ of clusters means that the total mass of iron
in the ICM is proportional to the total optical luminosity of the
cluster galaxies (Songaila, Cowie, \& Lilly 1990; Ciotti et al. 1991; Arnaud et
al. 1992; Renzini et al. 1993). The simplest interpretation 
is that the iron and all the metals now in the ICM have been
produced by the (massive) stars of the same stellar generation to
which belong the low-mass stars now radiating the bulk of the cluster
optical light.  Since much of the cluster light comes from
old spheroids (ellipticals and bulges), one can conclude
that {\it the bulk of cluster metals were produced by the stars
destined to make up the old spheroids that we see today in
clusters}.

It is also interesting to ask which galaxies have produced the bulk of
the iron and the other heavy elements, i.e. the relative contribution
as a function of the present-day luminosity of cluster galaxies. From
their luminosity function  it is easy to realize that
the bright galaxies (those with $L\, \gsim \, L^*$) produce the bulk of the
cluster light, while the dwarfs contribute a negligible amount of light, 
in spite of their dominating the galaxy counts by a large margin (Thomas
1999). In practice, most galaxies do not do much, while only the
brightest $\sim 3\%$ of all galaxies contribute $\sim 97\%$ of the
whole cluster light. Giants dominate the scene, while dwarfs do not 
count much.  Following the simplest interpretation, according
to which the metals were produced by the same stellar population that
now shines, one can also conclude that the bulk of the cluster metals
have been produced by the giant galaxies that contain most of the
stellar mass. The relative contribution of dwarfs to ICM metals may
have been somewhat larger than their small relative contribution to the cluster
light, since metals can more easily escape from their shallower
potential wells (Thomas 1999). Yet, this is unlikely to alter the
conclusion that the giants dominate metal production by a very large
margin.

\section{Metal Production: Type Ia vs. Type II Supernovae}

As is well known, clusters are now dominated by E/S0 galaxies, which
produce only Type Ia SNe at a rate of $\sim (0.16\pm 0.06)h_{70}^2$
SNU (Cappellaro, Evans, \& Turatto 1999), with 1 SNU corresponding to $10^{-12}$
SNe $\yr-1 L_{\rm B \odot}^{-1}$. Assuming such rate to have been
constant through cosmological times ($\sim$13 Gyr), the number of SNe~Ia
exploded in a cluster of present-day luminosity $\lb$ is therefore
$\sim 1.6\times 10^{-13}\times 1.3 \times 10^{10}\lb h_{70}^2\simeq
2\times 10^{-3}\lb$.  With each SN~Ia producing $\sim 0.7\,\msun$ of
iron, the resulting
\IMLR \ of clusters would be:
\begin{equation}
{\left(M^{\rm Fe}\over\lb\right)}_{\rm SN~Ia}\simeq 1.4\times 10^{-3}h_{70}^2,
\end{equation}
which falls short by a factor $\sim 10$ compared to the observed
cluster \IMLR \ ($0.015$ for $h_{70}=1$). The straightforward
conclusion is that either SNe~Ia did not play any significant role in
manufacturing iron in clusters, or their rate in what are now E/S0
galaxies had to be much higher in the past. This argues for a strong
evolution of the SN~Ia rate in E/S0 galaxies and bulges, with the past
average being $\sim 5-10$ times higher than the present rate (Ciotti
et al. 1991). This may soon be tested directly by observations.

In the case of SNe~Ia we believe to have a fairly precise knowledge of
the amount of iron released by each event, while the ambiguities
affecting the progenitors make theory unable to predict the evolution
of the SN~Ia rate past a burst of star formation (e.g., Greggio 1996).
The case of Type II SN's is quite the opposite: one believes to have
unambiguously identified the progenitors (stars more massive than
$\sim 8\, \msun$), while a great uncertainty affects the amount of iron
produced by each SN~II event as a function of progenitor's mass, 
$M^{\rm Fe}_{\rm II}(M)$. The SN luminosity at late times can be used to infer
the amount of radioactive Ni-Co (hence eventually iron) that was ejected,
and an early study indicated small variations from one event to another
($0.04-0.10\,\msun$; Patat et al. 1994). This led Renzini et al. (1993)
to assume $M^{\rm Fe}_{\rm II}$ to be a weak function of initial mass, with an 
average yield of $0.07\,\msun$ of iron per event (as in SN 1987A). 
More recent studies based on a larger sample of SN~II events 
have actually detected very large differences from one event to
another (ranging from $\sim 0.002\,\msun$ to $\sim 0.3\,\msun$; Turatto
2003). However, an average over 16 well-studied SNe~II gives 
$\langle M^{\rm Ni}\rangle = 0.062\,\msun$ (Hamuy 2003), close to the 
adopted value.

The total number of SNe~II, $N_{\rm II}$, is obtained by
integrating the stellar IMF from, for example, 8 to 100 $\msun$, with the IMF
being  $\psi(M)=3.0\lb\,M^{-(1+x)}$, where $\lb$ is the luminosity of the
stellar population when it ages to $\gsim 10^{10}$ yr (Renzini 1998b).
Clearly, the flatter the IMF slope the larger the number of massive stars per
unit present luminosity, the larger the number of SNe~II, and the 
larger the implied \IMLR. Thus, integrating the IMF one gets
\begin{equation}
{\left(M^{\rm Fe}\over\lb\right)}_{\rm SN~II}={M^{\rm Fe}_{\rm II}N_{\rm II}
         \over\lb}\simeq\cases{
    \hbox{0.003\quad{\rm for}$\; x=1.7$}\hfil\cr
    \hbox{0.009\quad{\rm for}$\; x=1.35$}\hfil\cr
    \hbox{0.035\quad{\rm for}$\; x=0.9.$}\hfil\cr}\
\end{equation}
Hence, if the Galactic IMF slope ($x=1.7$; Scalo
1986) applies also to cluster ellipticals, then SNe~II underproduce
iron by about a factor of $\sim 5$.
Instead, making all the observed iron by SNe~II would require an IMF 
flatter than Salpeter's $x=1.35$ (Renzini et al. 1993).

In summary, with an IMF with $1.35\, \lsim\, x\,\lsim \, 1.7$ and a past average rate
of SNe~Ia in ellipticals $\gsim 5$ times the present rate, 
the iron content of clusters and the global ICM
[$\alpha$/Fe] ratio are grossly accounted for, with SNe~Ia then having
produced $\sim$1/2--3/4 of the total cluster iron, not unlike in standard
chemical models of the Milky Way galaxy. This is not to say that this
has been firmly proved, but it seems to me to be premature to abandon
the attractive simplicity of a universal
nucleosynthesis process (i.e., IMF and SN~Ia/SN~II ratio) for embarking
toward more complex, multi-parametric scenarios.

\section{Metals from Galaxies to the ICM: Ejection vs. Extraction}

Having established that most metals in clusters are out of the parent
galaxies, it remains to be understood how they were transferred from
galaxies to the ICM.  There are two possibilities: extraction by ram
pressure stripping as galaxies plow through the ICM, and ejection by
galactic winds powered from inside galaxies themselves. In the latter
case the power can be supplied by SNe (the so-called star
formation feedback) and/or by AGN activity.
Three arguments favor ejection over extraction: 

\begin{itemize}

\item
There appears to
be no trend of either $\zfecm$ or the \IMLR \ with cluster temperature
or cluster velocity dispersion ($\sigma_{\rm v}$), while the
efficiency of ram pressure stripping should increase steeply with
increasing $\sigma_{\rm v}$.  

\item
Field ellipticals appear to be
virtually identical to cluster ellipticals. They follow basically the
same Mg$_2-\sigma$ relation (Bernardi et al. 1998, 2003), which does not
show any appreciable trend with the local density of galaxies. If 
stripping was responsible for
extracting metals from galaxies one would expect galaxies in low-density 
environments to have retained more metals, hence showing
higher metal indices for a given $\sigma$, which is not seen.

\item
Nongravitational energy injection of the ICM seems to be required to
account for the break of the self-similar X-ray luminosity-temperature
relation for groups and clusters (Ponman, Cannon, \& Navarro 1999; see
below). While galactic winds are an obvious vehicle for preheating,
no preheating would be associated with metal transfer by ram pressure.
\end{itemize}

One can quite safely conclude that metals in the ICM have been {\it
ejected} from galaxies by SN (or AGN) driven  winds,
rather than stripped by ram pressure (Renzini et al.
1993; Dupke \& White 1999). Two kinds of galactic winds are likely to
operate: {\it early winds} driven by the starburst forming much of the
galaxy's stellar mass itself, and {\it late winds} or outflows where
the gas comes from the cumulative stellar mass loss as the stellar populations
passively age.  Direct observational evidence for early winds
exists for Lyman-break galaxies (Pettini et al. 2001), as well as for
local massive starbursts (Heckman et al. 2000). Late winds are also
likely to operate, as the  stellar mass loss from the aging
population flows out of  spheroids, being either continuously
driven by a declining SN~Ia rate (Ciotti et al. 1991), or intermittently
by recurrent AGN activity (Ciotti \& Ostriker 2001).

\section{Metals as Tracers of ICM Preheating}

The total amount of iron in clusters represents a record of the overall
past SN activity as well as of the past mass and energy
ejected from cluster galaxies.  The empirical values of \IMLR \ can be used
to set a constraint on the energy injection into the ICM by SN-driven
galactic winds (Renzini
1994). The total SN heating is
given by the kinetic energy released by one SN ($\sim 10^{51}$ erg)
times the number of SNe that have exploded.  It is convenient to
express this energy per unit present optical light $\lb$, 
\begin{equation}
{E_{\rm SN}\over\lb}=10^{51}\,{ N_{\rm SN}\over\lb}=10^{51}\, 
\left({M^{\rm Fe}\over\lb}
\right)_{\rm tot}\,{1\over \langle M^{\rm Fe}\rangle}\simeq 10^{50}\quad 
({\rm erg}/\lsun), 
\end{equation}
where the  total (ICM+galaxies) \IMLR =0.015 $\msun/\lsun$ is adopted, and the
average iron release per SN event is assumed to be
$0.15\,\msun$ (appropriate if SNe~Ia and SNe~II
contribute equally to the iron production).
This estimate should be accurate to within a factor of 2 or 3.

The kinetic energy injected into the ICM by galactic winds, again per
unit cluster light, is given by 1/2 the ejected mass ($M^{\rm Fe}_{\rm ICM}
/Z^{\rm Fe}_{\rm w}$) times the square of the typical
wind velocity,
\begin{equation}
{E_{\rm w}\over\lb}={1\over 2} {M^{\rm Fe}_{\rm ICM}\over\lb}
{\langle \upsilon_{\rm w}^2\rangle \over \langle Z^{\rm Fe}_{\rm w}\rangle}\simeq 1.5\times
10^{49}{Z^{\rm Fe}_ \odot\over Z^{\rm Fe}_{\rm w}}\cdot
\left({\upsilon_{w}\over 500\,{\rm km}\,{\rm s}^{-1}}\right)^2\simeq
10^{49}\quad ({\rm erg}/\lsun),
\end{equation}
where the empirical \IMLR \ for the ICM has been used and the average
metallicity of the winds $Z^{\rm Fe}_{\rm w}$ is assumed to be 2 
times solar. As usual in the case of thermal winds, the wind velocity
$\upsilon_{\rm w}$ is of the order of the escape velocity from
individual galaxies. Again,
this estimate may be regarded as accurate to within a factor of 2 or so.

A first inference is that of order $\sim 5\%-20\%$ of the kinetic energy 
released by SNe is likely to survive as
kinetic energy of galactic winds, thus contributing to the heating of
the ICM.  A roughly similar amount goes into work to extract the gas
from the potential well of individual galaxies, while the rest of the
SN energy has to be radiated away locally and does not contribute to
the feedback.  This estimated energy injection represents a small
fraction of the thermal energy of the ICM of rich (hot) clusters and
so had only a minor impact  on the history of the ICM.  However,
in groups it represents a nonnegligible fraction of the thermal energy of
the ICM, thus affecting its evolution and present structure 
(Renzini 1994). The necessity of some nongravitational heating
(or preheating) was recognized from the observed break of the
self-similarity of the X-ray luminosity-temperature
relation, especially when groups are included (Ponman et al. 1999).

The estimated $\sim 10^{49}$ erg/\lsun \
correspond to a preheating of $\sim 0.1$ keV per particle, for a
typical cluster $M_{\rm ICM}/\lb\simeq 25\;\msun/\lsun$. This is
$\gsim 10$ times lower than the $\sim 1$ keV/particle preheating that
some models require to fit the cluster $L_{\rm X}-T$ relation (Wu, Fabian,
\& Nulsen 2000; Borgani et al. 2001, 2002; Tozzi \& Norman 2001; Pipino et
al. 2002; Finoguenov et al. 2003). This estimate depends somewhat on
the gas density (hence environment and redshift) where/when the energy
is injected, because what matters is the entropy change induced by the
preheating, $\Delta S=k\Delta T/n_{\rm e}^{2/3}$ (Kaiser 1991; Cavaliere, 
Colafrancesco, \& Menci 1993); hence, the  required energy decreases if it is 
injected at a lower gas density. Nevertheless, this extreme (1 keV/particle) 
requirement would
be met only if virtually all the SN energy were to go to increase the
thermal energy of the ICM. Such extreme preheating requirement
points toward an additional energy (entropy) source, such as AGN
energy injection (e.g., Valageas \& Silk 1999; Wu et al. 2000). Note,
however, that in powerful starbursts most SNe explode inside hot bubbles
made by previous SNe, thus reducing radiative losses, and the feedback
efficiency may approach unity (Heckman 2002).
More recently it has been suggested that preheating requirements may
be relaxed somewhat if the energy injection takes place at relatively
low density, so as to boost the entropy increase with less energy
deposition (Ponman, Sanderson, \& Finoguenov 2003). 
For example, preheating could take
place within the filaments, prior to their coalescing to form
clusters. Numerical simulations are exploring this possibility (see
the review by Evrard 2003), which may eventually reduce
the energetic requirement to be more in line with a conservative (i.e.,
$\sim 0.1-0.3$ keV/particle) SN-driven galactic wind scenario.

\section{Clusters vs. Field at $z=0$ and the Overall Metallicity of 
the Universe}

To what extent are clusters fair samples of the $z\sim 0$ Universe as
a whole? In many respects clusters look much different from the field,
for example in the morphological mix of galaxies, or in the star formation
activity, which in clusters has almost completely ceased while it is
still going on in the field. Yet, when we restrict ourselves to some
global properties clusters and field are not so different. For
example, the baryon fraction of the Universe is $\Omega_{\rm
b}/\Omega_{\rm m}\simeq 0.16\pm0.02$ (Bennett et al. 2003), which
compares to $\sim 0.15$ as estimated for clusters (White et al. 1993), 
adopting $h_{70}=1$. This tells us that no appreciable baryon vs. dark
matter segregation has taken place at a cluster scale (White et al. 1993).

Even more interesting may be the case of the stellar mass over baryon
mass in clusters and in the field. For the field in the local
Universe, Fukugita, Hogan, \& Peebles (1998)
 estimate $\Omega_*=0.0035h_{70}^{-1}$
for the stellar contribution to $\Omega$. From the 2dF $K$-band
luminosity function, Cole et al. (2001) estimate
$\Omega_*=0.0041h_{70}^{-1}$, with a $\sim 15\%$ uncertainty (adopting
a Salpeter IMF). The total baryon density is $\Omega_{\rm b}=0.039
h_{70}^{-2}$, as derived from standard Big Bang nucleosynthesis
(and confirmed by {\it WMAP}; Bennett et al. 2003). This gives a global
baryon to star conversion efficiency $\Omega_*/\Omega_{\rm b}\simeq
0.10h_{70}$ --- that is, over the whole cosmic time $\sim 10\%$ of the
baryons have been converted and locked into stars.
At the galaxy cluster level, the same efficiency can be measured directly,
and following White et al. (1993) one gets

\begin{equation}
{M_*\over\micm+M_*}\simeq {1\over 9.3h_{70}^{-3/2}+1}\simeq 0.1.
\end{equation} 
 
For clusters Fukugita et al. (1998) obtain  a $\sim 30\%$ larger
value, which, however, is well within the uncertainty
affecting these estimates. One can safely conclude that {\it the efficiency of 
baryon to galaxies/stars conversion has been $\sim 10\%$, quite the same in 
the ``field'' as well as within rich clusters of galaxies.} The environment
seems to be irrelevant!

Two interesting inferences can be drawn from this 
intriguing  cluster-field similarity: 

\begin{enumerate}

\item
{\it The metallicity of the present Universe is $\sim$1/3 solar.} 
The metallicity of the local
Universe has to be virtually identical to that measured in clusters
($\sim$1/3 solar), since star formation, and hence the ensuing metal
enrichment, have proceeded at the same level. In analogy to clusters,
a majority share of the metals now reside outside
galaxies in a warm intergalactic
medium (IGM) containing the majority of the baryons. Most baryons as well as
most metals in the local Universe remain unaccounted. 

\item
{\it The thermal energy (temperature) of the local Universe is about the same 
as the preheating energy of clusters.} Similar overall star formation activities
most likely result not only in similar metal productions but also in similar
energy deposition by galactic winds. Hence, the temperature in the local IGM
is likely to be $kT\approx 0.1-1$ keV, whatever the cluster preheating will
turn out to be.  Attempts are currently under way to detect this warm, 
metal-rich IGM. The detection of O~VI-absorbing clouds physically located 
within the Local Group is a first important step in this direction 
(Nicastro et al. 2003).

\end{enumerate}

\section{The Major Epoch of Metal Production} 

Most stars are either in spheroids or in disks. According to Fukugita
et al. (1998) $\sim$3/4 of the total mass in stars in the local
Universe is now in spheroids, $\sim$1/4 in disks, and less than 1\%
in irregular galaxies. Other authors give less extreme estimates; 
Dressler \& Gunn (1990) estimate that the stellar mass in spheroids and
disks is about the same (see also Benson, Frenk, \& Sharples 2002).  In
clusters the dominance of spheroids is likely to be even stronger than
in the general field.
The prevalence of spheroids offers an opportunity to estimate the epoch
(redshift) at which (most) metals were produced and disseminated,
since we now know quite well when most stars in spheroids were formed.

\subsection{In Clusters}

Following the first step in this direction (Bower, Lucey, \& Ellis 1992), I
believe that the most precise estimates of the age (redshift of
formation) of stellar populations in cluster elliptical galaxies come
from the tightness of several correlations, such as the
color-magnitude, fundamental plane, and the Mg$_2-\sigma$ relations,
and especially by such relations remaining tight all the way to $z\approx
1$ (Stanford, Eisenhardt, \& Dickinson 1998; van~Dokkum \& Franx 2001; see 
also Renzini 1999 for an extensive review and reference list). This has taught 
us that the best way of breaking the age-metallicity degeneracy is to
look back at high-redshift galaxies.  All evidence converges to
indicate that most stars in cluster ellipticals formed at $z\gsim 3$,
while only minor episodes of star formation may have occurred later.

With most of star formation having taken place at such high redshift,
most cluster metals should also have been produced and disseminated
at $z\gsim 3$. Little evolution of the ICM composition is then
expected all the way to high redshifts, with the possible exception of
iron from SNe~Ia, whose rate of release does not follow the star
formation rate, but is modulated by the distribution of the delays
between formation of the precursor and explosion time. Still, one expects
that the SN~Ia rate peaks shortly after a burst of star formation and
then rapidly declines, with most events taking place within 1--2 Gyr after
formation (e.g., Greggio \& Renzini 1983). If so, no appreciable evolution
of the iron abundance in clusters should be detectable from $z=0$ to
$z\approx 1$. Note, however, that {\it late winds} will keep enriching the ICM
at a decreasing rate approximately $\propto t^{-1.4}$ (Ciotti et al. 1991).

\subsection{In the Global Universe}

As already noted, at $z\approx 0$ field early-type galaxies show very little
differences with respect to their cluster analogs. 
Moreover, bulges appear very similar to ellipticals in integrated
properties, such as the Mg$_2-\sigma$ and fundamental plane relations
(Jablonka, Martin, \& Arimoto 1996; Falc\'on-Baroso, Peletier, \&
Balcells 2002). In the 
well studied case of the Milky Way bulge, no trace of stars younger
than halo-bulge globular clusters could be found (Zoccali et al. 2003).
At $z\approx 1$ old early-type galaxies are also found in sizable numbers in the
general field, 
although it appears that star formation may have been a little more extended 
than in clusters (Cimatti et al. 2002; Treu et al. 2002; Bell et al. 2003).

Therefore, spheroids in the general field appear almost as old as
cluster ellipticals, with the bulk of their stellar populations
having formed at $z\gsim 2-3$. In this spirit, Hogg et al. (2002) estimate that
at least 65\% of the stellar mass is at least 8 Gyr old, or formed at
$z>1$. With $\sim 50\%$ of the stellar mass
in spheroids that formed $\gsim 80\%$ of their mass at $z\gsim 2-3$,
one can conclude that $\gsim 30\%$ of the stellar mass we see today
was already in place by $z\approx 3$ (Renzini 1998a). This {\it indirect} 
estimate is $\sim 3$ times higher than {\it directly} measured
in the HDF-N (Dickinson et al. 2003). However, this latter result
may be subject to cosmic variance, given the small size of the explored
field, and indeed HDF-S appears to be much richer in massive galaxies, 
hence in stellar mass, at high redshift (Franx et al. 2003). 

\subsection{The Metallicity of the Universe at $z=3$}

With $\sim 30\%$ of all stars having formed by $z=3$, $\sim 30\%$
of the metals should also have been formed before such an early epoch.
I have argued that the global metallicity of the present-day Universe 
is $\sim$1/3 solar; hence, the metallicity of the $z=3$ Universe
should be $\sim$1/10 solar (Renzini 1998a). This simple argument supports the
notion of a {\it prompt initial enrichment} of the early Universe. While
$\sim 10\%$ solar at $z = 3$ is a very straightforward
estimate, observational tests are not so easy.

Figure 1.4 (adapted from Pettini 2003) shows that at
$z=3$ the Universe had developed extremely inhomogeneously in chemical
composition, with the metallicity ranging from supersolar in the
central regions of young/forming spheroids and in QSOs likely hosted
by them, down to $\sim 10^{-3}$ solar in the Ly$\alpha$ forest. Making
the proper (mass-) average abundance of the heavy elements requires one to
know the fractional mass of each baryonic component at $z=3$,
not an easy task.  While the Ly$\alpha$
forest may fill most of the volume at $z=3$ and perhaps contain most
of the baryons, it may contain as little as just a few percent of
the metals produced by $z=3$. At this early time most metals are
likely to be locked into stars, in metal-rich winds, and in shocked
IGM which has already diluted wind materials. This latter medium may
have been detected, thanks to its O~VI absorption (Simcoe, Sargent, \& 
Rauch 2002).

\begin{figure*}[t] 
\centering
\includegraphics[width=1.00\columnwidth,angle=0,clip]{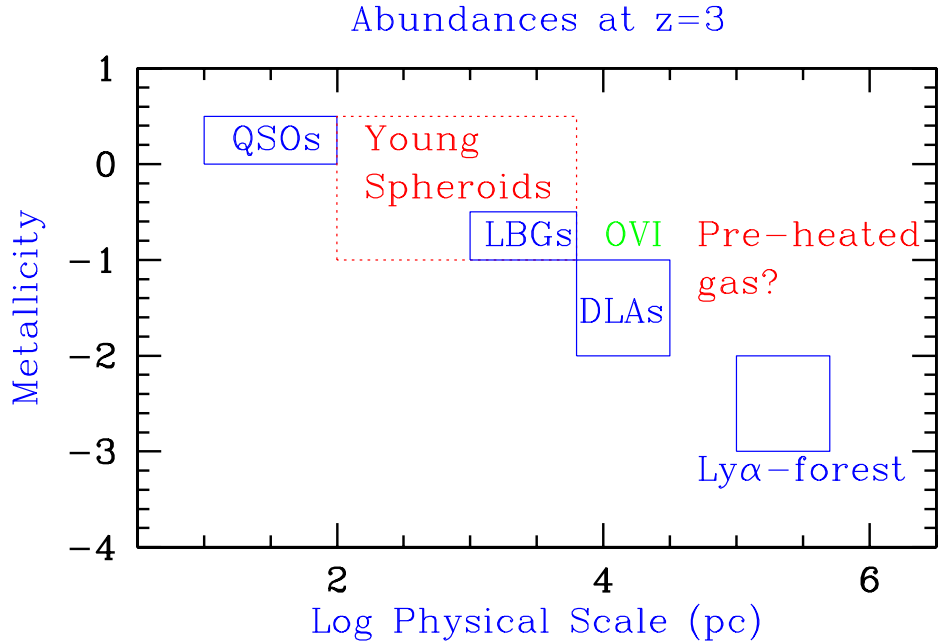}
\vskip 0pt \caption{
Summary of current knowledge of metal abundances at $z\approx 3$.
On the vertical axis the logarithmic abundance relative to solar is
reported. The horizontal axis gives the typical linear dimensions
of the structures for which direct abundance measurements are available.
This figure has been adapted from Pettini (2003) by the inclusion of
the box for ``young spheroidals,'' for which the estimate is indirect,
as based on the present-day observed metallicity range and on the
estimated redshift of formation. The figure also includes the approximate
location of the O~VI absorbers (Simcoe et al. 2002), and the hypothetical
location of the intergalactic medium enriched and preheated by early
galactic winds.}
\end{figure*}

\section{The Early Chemical Evolution of the Milky Way}

In the $K$ band the Galactic bulge luminosity is $\sim 
10^{10}\; L_{\rm K,\odot}$ (Kent, Dame, \& Fazio 1991), and in the $B$
band the bulge luminosity is $L^{\rm bulge}_{\rm B}\simeq 6\times 10^9
L_{\rm B, \odot}$. From
the empirical yield of metals in clusters ($\sim 0.2\times\lb\;
\msun)$, it follows that the Galactic bulge has produced $M_{\rm
Z}\simeq 0.2L^{\rm bulge}_{\rm B}=0.2\times 6\times 10^9\simeq
10^9\msun$ of metals. Where are all these metals? One billion solar masses of 
metals should not be easy to hide: part of it must be in the stars of the
bulge itself; part of it must have been ejected by winds. The stellar mass of 
the bulge follows from its $K$-band mass to light ratio, $M_*^{\rm
bulge}/L_{\rm K}=1$ (Kent 1992; Zoccali et al. 2003), and its
luminosity; hence, $M_*^{\rm bulge} \simeq 10^{10}\msun$. 
Its average metallicity is about solar or slightly lower
(McWilliam \& Rich 1994; Zoccali et al. 2003), i.e. $Z=0.02$, and
therefore the bulge stars altogether contain $\sim 2\times
10^8\msun$ of metals. Only $\sim$1/5 of the metals produced when the
bulge was actively star forming some 11--13 Gyr ago are still in the
bulge! Hence, $\sim 80\%$, or still $\sim 10^9\msun$, 
wad ejected into the surrounding space by an early wind.

At the time of bulge formation, such $\sim 10^9\msun$ of metals ran
into largely pristine ($Z=0$) material, experienced Rayleigh-Taylor 
instabilities leading to chaotic mixing, and established a distribution of
metallicities in a largely inhomogeneous IGM surrounding the young Milky
Way bulge. For example, this enormous amount of metals was able to bring to
a metallicity 1/10 solar (i.e., $Z=0.002$) about $5\times 10^{11}\msun$
of pristine material, several times the mass of the yet-to-be-formed
Galactic disk. Therefore, it is likely that the Galactic disk formed and grew
out of such pre-enriched material, which provides a quite natural solution
to the classical ``G dwarf problem'' (Renzini 2002). 

\section{Summary}

A number of interesting inferences are derived starting from a few
empirical facts, namely the iron and metal content of the ICM and
cluster galaxies, the fraction of the baryons locked into stars in
clusters and in the field, and the age and baryon fraction of stellar
populations of galactic spheroids. Such inferences include:

\begin{itemize}

\item
In clusters and in the general field alike there are more metals in the
diffused gas (ICM and IGM) than there are locked into stars inside
galaxies.  The loss of metals to the surrounding media is therefore a
major process in the chemical evolution of galaxies.

\item
In clusters and in the general field alike $\sim 10\%$ of the baryons are now
locked into stars inside galaxies. At this global level, the outcome of
star formation through cosmic time is largely independent of environment,
most likely just because a major fraction of all stars formed before cluster
formation.

\item
Various arguments support the notion that the metals now in the ICM/IGM
were ejected by galactic winds, rather then being extracted 
from galaxies by ram pressure.

\item
Having processed the same fraction of baryons into stars, the global
metallicity of the local Universe has to be nearly the same one can
measure in clusters, namely $\sim$1/3 solar.

\item
For the same reason, one expects the IGM to have experienced nearly the 
same amount of preheating as the ICM, and therefore   to be at
a temperature of $\sim 0.1-1$ keV, whatever is the amount of preheating
that is required for clusters.
 
\item
Given the predominance and formation redshift of galactic spheroids,
both in clusters as well as globally in the Universe, it is likely
that the Universe experienced a prompt metal enrichment, with the
global metallicity possibly reaching $\sim$1/10 solar already at $z\approx 3$. 
However, most metals remain unaccounted both at low as well as high redshift,
and may reside in a warm IGM, the existence  of which we have only preliminary
observational hints.

\item
This same scenario may well hold down to the scale of our own Milky Way galaxy,
with early winds from the forming Galactic bulge having pre-enriched to
$\sim$1/10 solar a much greater mass of gas, out of which the Galactic disk
started to form and grew.

\end{itemize}


\begin{thereferences}{}

\bibitem{}
Aguirre, A., Hernquist, L., Schaye, J., Katz, N., Weinberg, D. H., \& 
Gardner, J. 2001, ApJ, 561, 521 

\bibitem{}
Anders, E., \& Grevesse, N. 1989, Geochimica et Cosmochimica Acta, 53, 197

\bibitem{}
Arimoto, N., Matsushita, K., Ishimaru, Y., Ohashi, T., \& Renzini, A. 1997, 
ApJ, 477, 128

\bibitem{}
Arnaboldi, M., et al. 2003, AJ, 125, 514

\bibitem{}
Arnaud, M., \& Evrard, A. E. 1999, MNRAS, 305, 631

\bibitem{}
Arnaud, M., Rothenflug, R., Boulade, O., Vigroux, R., \& Vangioni-Flam, E. 
1992, A\&A, 254, 49

\bibitem{} 
Bell, E. F., Wolf, C., Meisenheimer, K., Rix, H.-W., Borch, A., Dye, S., 
Kleinheinrich, M., \& McIntosh, D. H. 2003, ApJ, submitted (astro-ph/0303394)

\bibitem{} 
Bennett, C. L., et al. 2003, ApJ, submitted (astro-ph/0302207)

\bibitem{}
Benson, A. J., Frenk, C. S., \& Sharples, R. M. 2002, ApJ, 574, 104

\bibitem{} 
Bernardi, M., et al. 2003, AJ, 125, 1882

\bibitem{} 
Bernardi, M., Renzini, A., da Costa, L.~N., Wegner, G., Alonso, M.~V.,
Pellegrini, P.~S., Rit\'e, C., \& Willmer, C.~N.~A. 1998, \apj, 508, L143

\bibitem{}
Borgani, S., Governato, F., Wadsley, J., Menci, N., Tozzi, P., Lake,
G., Quinn, T., \& Stadel, J. 2001, ApJ, 559, L71

\bibitem{}
Borgani, S., Governato, F., Wadsley, J., Menci, N., Tozzi, P., 
Quinn, T., Stadel, J., \& Lake, G. 2002, MNRAS, 336, 424

\bibitem{}
Bower, R. G., Lucey, J. R., \& Ellis, R. S. 1992, MNRAS, 254, 613

\bibitem{}
Buote, D. A. 2000, MNRAS, 311, 176

\bibitem{}
Cappellaro, E., Evans, R., \& Turatto, M. 1999, A\&A, 351, 459

\bibitem{}
Cavaliere, A., Colafrancesco, S., \& Menci, N. 1993, ApJ, 415, 50

\bibitem{}
Chiosi, C. 2000, A\&A, 364, 423

\bibitem{}
Cimatti, A., et al. 2002, A\&A, 381, L68

\bibitem{}
Ciotti, L., D'Ercole, A., Pellegrini, S., \& Renzini, A. 1991, ApJ, 376, 380

\bibitem{}
Ciotti, L., \& Ostriker, J.P. 2001, ApJ, 551, 131

\bibitem{}
Cole, S., et al. 2001, MNRAS, 326, 255

\bibitem{}
David, L. P., Jones, C., Forman, W., \& Daines, S. 1994, \apj, 428, 544 

\bibitem{}
Davis, D. S., Mulchaey, J. S., \& Mushotzky, R. F. 1999, ApJ, 511, 34

\bibitem{}
De Grandi, S., \& Molendi, S. 2001, ApJ, 551, 153

\bibitem{}
------. 2002, in Chemical Enrichment of Intracluster and Intergalactic Medium, 
ed. R. Fusco-Femiano \& F. Matteucci (San Francisco: ASP), 3

\bibitem{}
Dickinson, M., Papovich, C., Ferguson, H. C., \& Budavari, T. 2003, 
ApJ, 587, 25 

\bibitem{}
Dressler, A., \& Gunn, J. E. 1990, in Evolution of the Universe of Galaxies, 
Proceedings of the Edwin Hubble Centennial Symposium, ed. R. G. Kron 
(San Francisco: ASP), 200

\bibitem{} 
Dupke, R. A., \& White, R. E., III 2000, ApJ, 537, 123

\bibitem{} 
Evrard, A. E. 2003, in Carnegie Observatories Astrophysics Series, Vol. 3:
Clusters of Galaxies: Probes of Cosmological Structure and Galaxy Evolution,
ed. J. S. Mulchaey, A. Dressler, \& A. Oemler (Cambridge: Cambridge Univ.
Press), in press

\bibitem{}
Falc\'on-Barroso, J., Peletier, R. F., \& Balcells, M. 2002, MNRAS, 335, 741

\bibitem{}
Ferguson, H. C., Tanvir, N. R., \& von Hippel, T. 1998, Nature, 391, 461

\bibitem{}
Finoguenov, A., Arnaud, M., \& David, L. P. 2001, ApJ, 555, 191 

\bibitem{}
Finoguenov, A., Borgani, S., Tornatore, L., \& B\"ohringer, H. 2003, A\&A, 
398, L35

\bibitem{}
Finoguenov, A., David, L. P., \& Ponman, T. J. 2000, ApJ, 544, 188

\bibitem{}
Finoguenov, A., Jones, C.,  B\"ohringer, H., \& Ponman, T.J. 2002, ApJ, 578, 74 

\bibitem{}
Franx, M., et al. 2003, \apj, 587, L79

\bibitem{}
Fukazawa, Y, Ohashi, T., Fabian, A. C., Canizares, C. R., Ikebe, 
Y., Makishima, K., Mushotzky, R. F., \& Yamashita, K. 1994, PASJ, 46, L55

\bibitem{}
Fukugita, M., Hogan, C. J., \& Peebles, P. J. E. 1998, ApJ, 503, 518

\bibitem{}
Gibson, B., \& Matteucci, F. 1997, MNRAS, 291, L8

\bibitem{}
Greggio. L. 1996, in The Interplay between Massive Star Formation, the ISM 
and Galaxy Evolution, ed. D. Kunth et al. (Gif-sur-Yvettes: Edition 
Fronti\`eres), 89

\bibitem{}
Greggio. L., \& Renzini, A. 1983, A\&A, 118, 217

\bibitem{}
Hogg, D. W., et al. 2002, AJ, 124, 646

\bibitem{}
Hamuy, M. 2003, in Core Collapse of Massive Stars, ed. C. L. Fryer (Dordrecht: 
Kluwer), in press (astro-ph/0301006)

\bibitem{}
Heckman, T. M. 2002, in Extragalactic Gas at Low Redshift, ed. J. S. Mulchaey 
\& J. Stocke (San Francisco: ASP), 292

\bibitem{}
Heckman, T. M., Lehnert, M. D., Strickland, D. K., \& Armus, L. 2000, ApJS, 
129, 493

\bibitem{}
Ishimaru, Y., \& Arimoto, N. 1997, PASJ, 49, 1

\bibitem{}
Jablonka, P., Martin, P., \& Arimoto, N. 1996, AJ, 112, 1415

\bibitem{}
Kaiser, N. 1991, ApJ, 383, 104

\bibitem{}
Kent, S. M. 1992, \apj, 387, 181

\bibitem{}
Kent, S. M., Dame, T. M., \& Fazio, G. 1991, \apj, 378, 131

\bibitem{}
Larson, R. B., \& Dinerstein, H. L. 1975, PASP, 87, 511

\bibitem{}
Loewenstein, M. 2001, ApJ, 557, 573

\bibitem{}
Loewenstein, M., \& Mushotzky, R. F. 1996, ApJ, 466, 695

\bibitem{}
Maraston, C., Greggio, L., Renzini, A., Ortolani, S., Saglia, R. P.,
Puzia, T. H., \& Kissler-Patig, M. 2003, A\&A, 400, 823

\bibitem{}
Matteucci, F., \& Vettolani, G. 1988, A\&A, 202, 21

\bibitem{}
McWilliam, A., \& Rich, R. M. 1994, \apjs, 91, 794

\bibitem{}
Mitchell, R. J., Culhane, J. L., Davison, P. J., \& Ives, J. C. 1976, MNRAS, 
175, 29P

\bibitem{}
Mohr, J. J., Mathiesen, B., \& Evrard, A. E. 1999, ApJ, 627, 649

\bibitem{}
Mulchaey, J. S., Davis, D. S., Mushotsky, R. F., Burstein, D. 1993, \apj, 
404, L9

\bibitem{}
------. 1996, ApJ, 456, 80

\bibitem{}
Mushotzky, R. F. 1994, in Clusters of Galaxies, ed. F. Durret, A. Mazure, \& 
J. T. Thanh Van (Gyf-sur-Yvette: Editions Fronti\`eres), 167

\bibitem{}
------. 2002, Phil. Trans. R. Soc. Lond. A, 360, 2019

\bibitem{}
Mushotzky, R. F., Loewenstein, M., Arnaud, K. A., Tamura, T., Fukazawa, Y., 
Matsushita, K., Kikuchi, K., \& Hatsukade, I. 1996, ApJ, 466, 686

\bibitem{} 
Nicastro, F., et al. 2003, Nature, 421, 719

\bibitem{}
Patat, F., Barbon, R., Cappellaro, E., \& Turatto, M. 1994, A\&A, 282, 731

\bibitem{}
Pettini, M. 2003, in Cosmochemistry: The Melting Pot of Elements (Cambridge:
Cambridge Univ. Press), in press (astro-ph/0303272)

\bibitem{}
Pettini, M., Shapley, A.~E., Steidel, C.~C., Cuby, J.-G., Dickinson, M.,
Moorwood, A.~F.~M., Adelberger, K.~L., \& Giavalisco, M. 2001, \apj, 554, 981

\bibitem{}
Pipino, A., Matteucci, F., Borgani, S., \& Biviano, A. 2002, NewA, 7, 227

\bibitem{}
Ponman, T. J., Allan, D. J., Jones, L. R., Merrifield, M., McHardy, I. M.,  
Lehto, H. J., \& Luppino, G. A. 1994, Nature, 369, 462 

\bibitem{}
Ponman, T. J., Cannon, D. G., \& Navarro, J. F. 1999, Nature, 397, 135

\bibitem{}
Ponman, T. J., Sanderson, A. J. R., \& Finoguenov, A. 2003, MNRAS, in press 
(astro-ph/0304048)

\bibitem{}
Pratt, G. W., \& Arnaud, M. 2003, A\&A, in press (astro-ph/0304017)

\bibitem{}
Renzini, A. 1994, in Clusters of Galaxies, ed. F. Durret, A. Mazure, \&
J. T. Thanh Van (Gyf-sur-Yvette: Editions Fronti\`eres), 221

\bibitem{}
------. 1997, ApJ, 488, 35

\bibitem{}
------. 1998a, in The Young Universe: Galaxy Formation and Evolution at 
Intermediate and High Redshift, ed. S. D'Odorico, A. Fontana, \& E. Giallongo 
(San Francisco: ASP), 298

\bibitem{}
------. 1998b, AJ, 115, 2459

\bibitem{}
------. 1999, in The Formation of Galactic Bulges, ed. C.~M.  Carollo, H.~C. 
Ferguson, \& R.~F.~G. Wyse (Cambridge: Cambridge Univ. Press), 9

\bibitem{}
------. 2000, in Large Scale Structure in the X-ray Universe, ed. M.
Plionis \& I. Georgantopoulos (Paris: Atlantisciences), 103

\bibitem{}
------. 2002, in Chemical Enrichment of Intracluster and Intergalactic Medium, 
ed. R. Fusco-Femiano \& F. Matteucci (San Francisco: ASP), 331

\bibitem{}
Renzini, A., Ciotti, L., D'Ercole, A., \& Pellegrini, S. 1993, ApJ, 419, 52 

\bibitem{}
Scalo, J. M. 1986, Fund. Cosmic Phys. 11, 1

\bibitem{}
Simcoe, R. A., Sargent, W. L. W., \& Rauch M. 2002, ApJ, 578, 737

\bibitem{}
Songaila, A., Cowie, L. L., \& Lilly, S. J. 1990, ApJ, 348, 371

\bibitem{}
Stanford, S. A., Eisenhardt, P. R., \& Dickinson, M. 1998, ApJ, 492, 461

\bibitem{}
Thomas, D. 1999, in Chemical Evolution from Zero to High Redshift, ed. J. R. 
Walsh \& M. R. Rosa (Berlin: Springer), 197

\bibitem{}
Tozzi, P., \& Norman, C. 2001, ApJ, 546, 63

\bibitem{}
Treu, T., Stiavelli, M., Casertano, S., M{\o}ller, P., Bertin, G. 2002, Apj, 
564, L13

\bibitem{}
Tsuru, T. 1992, Ph.D. Thesis, Univ. Tokyo 

\bibitem{}
Turatto, M. 2003, in Supernovae and Gamma-Ray Bursters, ed. K. W. Weiler
(Berlin: Springer), in press (astro-ph/0301107)

\bibitem{}
Valageas, P., \& Silk, J. 1999, A\&A, 350, 725

\bibitem{}
van Dokkum, P. G., \& Franx, M. 2001, ApJ, 553, 90

\bibitem{}
Vigroux, L. 1977, A\&A, 56, 473

\bibitem{}
White, D. A. 2000, MNRAS, 312, 663

\bibitem{} 
White, S. D. M., Navarro, J. F., Evrard, A. E., \& Frenk, C. S. 1993, Nature, 
366, 429

\bibitem{}
Wu, K. K. S., Fabian, A. C., \& Nulsen, P. E. J. 2000, MNRAS, 318, 889 

\bibitem{}
Zoccali, M., et al. 2003, A\&A, 399, 931

\end{thereferences}
\end{document}